\documentclass[a4paper,12pt]{article}
\pdfoutput=1 

\usepackage{jheppub_X} 
                     
\usepackage{booktabs,colortbl}
\usepackage[capitalise]{cleveref}				
\usepackage[alsoload=hep]{siunitx}		
\usepackage{cancel,xcolor,multirow}
\usepackage[normal,font=small]{caption}
\usepackage{subcaption}
\usepackage{slashed}
\usepackage{bm}
\usepackage{mathtools}					
\usepackage{tikz}						
\usepackage{comment}
\usepackage[bottom]{footmisc}			

\providecommand{\ord}{O}
\providecommand{\ie}{\emph{i.e.}}
\providecommand{\eg}{\emph{e.g.}}
\DeclareSIUnit{\keV}{\kilo\electronvolt}
\DeclareSIUnit{\MeV}{\mega\electronvolt}
\DeclareSIUnit{\GeV}{\giga\electronvolt}
\DeclareSIUnit{\TeV}{\tera\electronvolt}
\DeclareSIUnit{\nb}{\nano\barn}
\DeclareSIUnit{\pb}{\pico\barn}
\DeclareSIUnit{\year}{yr}

\renewcommand{\vec}[1]{\ensuremath{\mathbf{#1}}} 

\newcommand{\avg}[1]{\left\langle #1 \right\rangle} 

\title{
(Machine) Learning to Do More with Less
}

\author{Timothy Cohen,}
\author{Marat Freytsis,}
\author{and Bryan Ostdiek}

\emailAdd{tcohen@uoregon.edu}
\emailAdd{freytsis@uoregon.edu}
\emailAdd{bostdiek@uoregon.edu}

\affiliation{Institute of Theoretical Science, University of Oregon, Eugene, Oregon 97403, USA}

\abstract{Determining the best method for training a machine learning algorithm is critical to maximizing its ability to classify data. In this paper, we compare the standard ``fully supervised" approach (which relies on knowledge of event-by-event truth-level labels) with a recent proposal that instead utilizes class ratios as the only discriminating information provided during training.  This so-called ``weakly supervised" technique has access to less information than the fully supervised method and yet is still able to yield impressive discriminating power.  In addition, weak supervision seems particularly well suited to particle physics since quantum mechanics is incompatible with the notion of mapping an individual event onto any single Feynman diagram. We examine the technique in detail -- both analytically and numerically --  with a focus on the robustness to issues of mischaracterizing the training samples.  Weakly supervised networks turn out to be remarkably insensitive to a class of systematic mismodeling. Furthermore, we demonstrate that the event level outputs for weakly versus fully supervised networks are probing different kinematics, even though the numerical quality metrics are essentially identical. This implies that it should be possible to improve the overall classification ability by combining the output from the two types of networks. For concreteness, we apply this technology to a signature of beyond the Standard Model physics to demonstrate that all these impressive features continue to hold in a scenario of relevance to the LHC. Example code is provided on \href{https://github.com/bostdiek/PublicWeaklySupervised/tree/master}{GitHub}.}

\begin{document} 
\maketitle
\flushbottom

\setcounter{page}{2}
\section{Introduction}
\label{sec:intro}

Machine learning currently stands as one of the most exciting fields in computer science. The basic premise is to develop tools that can classify multi-variate data into categories by examining the data itself. Applications of these powerful tools within the fundamental physics community have been steadily gaining traction. Collaborations at the LHC have been incorporating this technology into various aspects of their characterization of objects, \eg, bottom~\cite{Aad:2015ydr,CMS:2016kkf} and charm~\cite{ATL-PHYS-PUB-2015-001, CMS:2016knj} quark tagging.  Applications in the field of jet substructure~\cite{Cogan:2014oua, Almeida:2015jua, deOliveira:2015xxd, Baldi:2016fql, Guest:2016iqz, Datta:2017rhs, Shimmin:2017mfk} are quickly maturing as well. Techniques such as deep learning, image recognition, adversarial networks, and symbolic regression have also found their way into the high energy literature \cite{Cranmer:2004gna, WHITESON20091203, Cogan:2014oua, Baldi:2014kfa,Almeida:2015jua, deOliveira:2015xxd,Searcy:2015apa,Baldi:2016fzo,Komiske:2016rsd, Barnard:2016qma, Pang:2016vdc, Kasieczka:2017nvn, Louppe:2016ylz, deOliveira:2017pjk, Shimmin:2017mfk, Pearkes:2017hku, Louppe:2017ipp, Huffman:2017vwp, He:2017aed}.

An outstanding problem for the reliable interpretation of machine learning outputs for physics problems has been that of how to gain control over the propagation of uncertainties of the input data through the learning algorithms. Some aspects of this concern can be addressed by emphasizing data-driven techniques so that as little of the machine learning architecture as possible depends on our (necessarily incomplete) models of fundamental interactions. Others can be aided through the use of adversarial networks to ensure that the learning step does not overly focus on a poorly modeled or correlated feature~\cite{Shimmin:2017mfk,deOliveira:2017pjk}. Often, the approach is to systematically vary properties of the input data, and use the spread in the output measurement to determine an error profile. However, in many cases the labels that come with the input data in order to train the algorithms are saddled with their own uncertainties that are correlated in a complicated way with the data itself. And if a confidence interval needs to be assigned to a given output of the classification, the problem becomes harder still.

With these questions in mind, the goal of this work is to explore the properties of \emph{weakly supervised neural networks}~\cite{Dery:2017fap}.  Machine learning algorithms learn to distinguish between signal and background by training on some user specified input data. This step is formulated so that the input data sets are only labeled by a measure of the fraction of signal events contained in the set. In the computer science literature, this class of problems goes by the name Learning from Label Proportions~\cite{DBLP:journals/jmlr/QuadriantoSCL09,2014arXiv1402.5902Y}.\footnote{This approach is closely related to the older ``bag label''  or ``multiple-instance'' formalism more extensively studied in the computer science literature~\cite{DIETTERICH199731,AMORES201381}. This is a variation on supervised learning where individual labels are not known, but in a ``bag'' of samples, the number of positive samples may be known.}  This is in contrast with the more widespread \emph{fully supervised neural network}, where one of the training sets is labeled as pure signal and the other as pure background. It opens the possibility of extending data-driven techniques into the realm of supervised learning methods, as weakly supervised networks could be trained on real data where absolute purity is unachievable.

In our view, weak supervision as a formalism touches on a much deeper connection to the nature of observables in particle physics.  Quantum mechanics implies that no single event can ever be mapped onto an individual Feynman diagram at the fundamental level --- event by event, the distinction between signal and background is not a well-defined notion.  More practically speaking, isolating interesting sets of LHC data will always require some preselection.  Assuming the selection is made such that there will be a non-trivial signal efficiency, the bin will be a mix of signal and background events, making the connection with weakly supervised training obvious.

Given this motivation, it becomes interesting to explore weak supervision beyond the compelling statements made in~\cite{Dery:2017fap}.  Therein, this technique was applied to the question of quark--gluon discrimination, demonstrating that weak supervision appears to perform just as well as fully supervised neural networks while providing a way to potentially avoid concerns about the imperfect modeling of quark and gluon jets in (simulated) fully-labeled data. However, the need for fractional labels derived from somewhere only moves concerns about modeling uncertainties elsewhere. Although overall sample fractions are indeed more theoretically robust than detailed event-by-event modeling \cite{Badger:2016bpw}, they do still come with their own uncertainties, whose effect on the learning algorithms one would like to be able to track. 

One of the main points of this work is to demonstrate that, in fact, weak supervision is robust to such uncertainties and systematic errors.  As we show below, the machine learning algorithm is smart enough to learn to distinguish signal from background when training on mixed samples, and that the performance is almost independent of the ``correctness'' of the fraction labels.  We demonstrate this surprising statement numerically. We also explore this behavior analytically.

In particular, this observation broadens the appeal of weak supervision to questions where data-driven training is less viable, like beyond the Standard Model (BSM) searches, on which there has been less focus on applying machine learning techniques. (Although see \eg,~\cite{Ellis:2012sd,Ellis:2012zp, Baldi:2014kfa,Cohen:2016nzv, Iwamoto:2017ytj, Barello:2016zlb, Buckley:2011kc, Bornhauser:2013aya, Caron:2016hib,Bertone:2016mdy,Bechtle:2017vyu, Metodiev:2017vrx} for a few exceptions.) With no BSM physics yet found at the LHC, there is no real-world data to train classifiers of BSM physics. However, concerns about propagation of uncertainties through the ``black box'' of machine learning are still present. Moreover, concerns about how accurately the Standard Model (SM) backgrounds are being modeled remain. We will demonstrate that in such a BSM scenario, for the reasons outlined above, weakly supervised classification still provides advantages.

The rest of this paper is organized as follows. \Cref{sec:review} provides a review of neural networks generally, along with a discussion of both weakly and fully supervised training.  \Cref{sec:toys} applies this technology to a toy model in order to compare the performance of weakly and fully supervised networks and to demonstrate the surprising robustness of weak supervision.  \Cref{sec:Analytics} gives an analytic argument for this insensitivity to mislabeling.  \Cref{sec:BSM} applies these tools to a beyond the Standard Model example by comparing the signal of gluinos to $Z + \text{jets}$.  Finally, \Cref{sec:disc} provides a demonstration that weak and fully supervised networks probe complementary regions of phase space, and closes with a discussion of many future directions. A discussion of how different loss functions affect the classifier performance is contained in \cref{Sec:Appendix}.

\section{Review of Weak Versus Full Supervision}
\label{sec:review}
Learning from Label Proportions~\cite{DBLP:journals/jmlr/QuadriantoSCL09,2014arXiv1402.5902Y} concerns itself with a class of supervised machine learning problems wherein only the aggregate properties of sets are presented for training, while definite properties of individual members are unknown to the algorithm. While it has already seen application in more traditional machine learning domains, the use of these techniques in the context of high energy physics was only recently presented under the moniker of weakly supervised classification. 
In \cite{Metodiev:2017vrx},  learning from label proportions is treated as one implementation of the paradigm of weak supervision.

For physics applications, one appeal is that since truth-level labels are not required, the networks can be trained on real, as opposed to simulated, data.  The authors of~\cite{Dery:2017fap} showed that weakly supervised classifiers as trained on distorted data demonstrated lower performance degradation than fully supervised ones.  This motivates a more complete characterization of the error tolerance of weak supervision, and in doing so we find some surprising behavior and resultant additional applications. But first, in order to orient the reader, we review traditional fully supervised approaches to event classification and contrast them with the newer weakly supervised method.

\begin{figure}[tb]
	\begin{center}
		\includegraphics[width=\columnwidth]{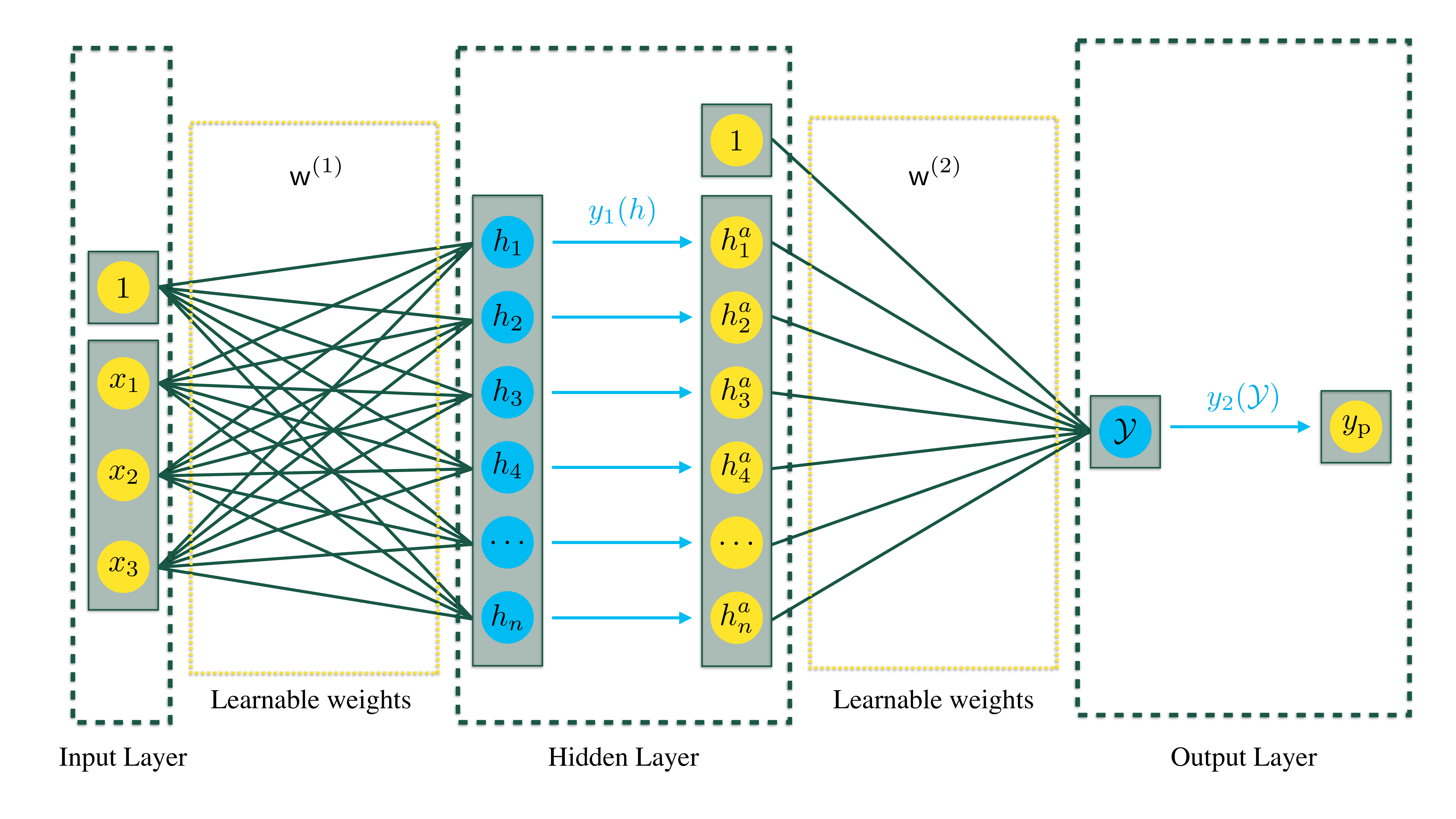}
		\caption{A schematic representation of a network with three layers (one hidden), such as the networks used in this paper. Each of the layers is marked by a dashed green box. All of the yellow circles represent a real number (node). The green lines are the connections between nodes and are each learnable parameters. The light blue arrows map the node activity (number in blue) to the node by applying the activation function.}
		\label{fig:NNRep}
	\end{center}
\end{figure}

Throughout this paper, we will be examining binary classification between events of type 0 and 1 using an artificial neural network.  A schematic of an representative network is shown in \cref{fig:NNRep}.  Artificial neural networks are comprised of a set of layers, denoted by the dashed green boxes.  In the figure, and in all of the results in this paper, the network consists of three layers: an input layer, a hidden layer, and the output layer.

Each layer contains a fixed number of nodes $n$ (sometimes referred to as neurons), denoted by the yellow circles in the figure.  The input layer contains a node for each feature, \emph{i.e.}, independent variable, in the input data; $n_\text{input}$ is fixed by the input data set. It is customary to add a constant \emph{bias} node to this layer as well, which can be thought of as the analog of adding a $y$-intercept when fitting data with a line.  For the hidden layer(s), there is the freedom to choose $n$ for each layer.  This is an example of a user-defined parameter or \emph{hyper-parameter}, to be distinguished from the fitting parameters or \emph{weights}.   We follow the standard method and add a bias node to the hidden layer as well. The nodes of the input and hidden layer (including the bias nodes) can be viewed as two vectors, with dimension $n_{\text{inputs}} +1$ and $n_{\text{hidden}} +1$, respectively.  Last, the output layer contains only one node, and the value it yields is the prediction of the network.

To pass the input signal to the output, the nodes of the layers need to be connected.  The connections are marked by the green lines in \cref{fig:NNRep}, such that transferring from one layer to the next is simply matrix multiplication.  We have represented these matrices in the sketch by the dotted yellow boxes, with labels $\mathsf{w}^{(1,2)}$.  Thus, the hidden layer is connected to the input layer by $h_i = w^{(1)}_{ij} x_{j}$.  The dimension of $\mathsf{w}^{(1)}$ is $ n_{\text{hidden}} \times (n_{\text{input}}+1)$, and each element of the matrix is a learnable weight.  Similarly, the output layer is connected to the hidden layer via the matrix $\mathsf{w}^{(2)}$, which contains $1 \times (n_{\text{hidden}}+1)$ learnable weights, giving $\mathcal{Y}_i = w^{(2)}_{ij} h^a_j$. 

In the figure, the vectors $\vec{h}$ and $\vec{\mathcal{Y}}$ are marked by blue circles, to represent the \emph{activity} of the nodes.  The activity of each node is then passed through an activation function, which is marked as the blue arrows taking $\vec{h} \rightarrow \vec{h}^a$ and $\vec{\mathcal{Y}}\rightarrow y_{\text{p}}$. For all of the networks considered in this paper, we use the activation function
\begin{equation}
\label{eqn:Sigmoid}
  y(h) = \frac{1}{1+e^{-h}}
       = \tikz[baseline=1pc,domain=-1:1]{	
           \draw[color=gray,<->] (-1.04,0) -- (1.04,0);
           \draw[color=gray,->] (0,0) -- (0,1.04);
           \draw[color=green!50!black, thick] plot[domain=-1:1,samples=50]
             ({\x}, {1/(1+exp(-9*\x))});}\,,
\end{equation}
also known as the logistic or sigmoid function: it maps $\mathbb{R} \rightarrow (0,1)$ continuously, with a shape that mimics that of a step function. In general, the choice of the activation function is another hyper-parameter. The neuron can then interpolate between on ($\equiv 1 \equiv \text{signal}$) or off ($\equiv 0 \equiv \text{background}$) depending on if the activity is either positive or negative. The sigmoid function is useful because it is smooth, continuous, and has a trivial derivative, which all contribute to efficient network training.

A few common generalizations can extend the flexibility of a neural network. For \emph{deep learning} extra hidden layers could be added, with the number of hidden layers and the number of nodes in each layer being free hyper-parameters. Also, while we use \cref{eqn:Sigmoid} for each of our activation steps, a wide variety of different function can be used on a node-by-node basis; it is not even necessary that they be deterministic.

To summarize, for a binary classification problem the output of the network $y_p= g(\vec{x};\theta)$ for a given event can be generalized to a function of the event features \vec{x} and the learnable parameters $\theta = \{\mathsf{w}^{(1,2)}\}$.  For brevity we will not write out the full expression for $y_p$ which can be inferred from the structure of \cref{fig:NNRep}.  Once the network is trained, $\theta$ becomes fixed, and an event will therefore always yield the same prediction. On the other hand, during training the goal is to find a well-performing set of parameters $\theta$ by minimizing a \emph{loss function}, which compares the known labels of the training set with the network output as we will detail below. The loss function is minimized by taking its gradient with respect to the parameters in $\theta$, for fixed $\vec{x}$. As a technical aside, the initial weights of $\mathsf{w}^{(1,2)}$ before training are chosen from a normal distribution. The specification of this initial condition is yet another hyper-parameter.

A variety of loss functions have been proposed in the literature. For example, in linear regression problems, where one wants to predict a number and not a class, a common loss function is the mean squared error, determined by how far off the predicted value is from the true value. The weights can then be tuned to minimize the error.\footnote{As with all machine learning, there is always a worry that one \emph{overtrains} such that the neural network becomes overly sensitive to the detailed properties of the training data set because of the large number of learnable parameters contained in $\theta$. In order to mitigate this issue, we use an independent validation data set which is piped through the network at the end of training.  Then we check that the loss function returns values that are within tolerance of the loss function outputs that are achieved for the training data.}

The approach in classification problems is similar, but instead of predicting a number, the goal is to predict the class. As shown above, the last step of our networks is to pass the output activity, $\vec{\mathcal{Y}}$, through the activation function. The larger $\vec{\mathcal{Y}}$ is, the closer to 1 the prediction. Conversely, large negative values of $\vec{\mathcal{Y}}$ map to a prediction of class 0. The absolute size of $\vec{\mathcal{Y}}$ can be thought of as how confident the network is about the prediction. This highlights why, while one could still use the mean squared error as the loss function, it is not ideal for classification. For instance, imagine comparing two events with truth labels $y_{t,i} = 0$, where one has an output activity of $\vec{\mathcal{Y}} = 3 \rightarrow y_{p}\sim0.95$ and the other has $\vec{\mathcal{Y}} = 6 \rightarrow y_p \sim 0.998$. The error for the second one is only about $5\percent$ larger, even though the network was twice as `confident' in its wrong assumption that the event was from class 1. The standard approach to classification uses a loss function which penalizes not just having the wrong answer, but also the level of confidence in the prediction.

For the networks used in the main body of this paper, we implement the binary cross entropy (BCE) loss function for both the fully and weakly supervised networks,
\begin{equation}
\label{eqn:FullLoss}
  \ell_\text{BCE} \big(\{y_t\}, \{y_p\} \big)
    = \sum_{i\, \in\, \text{samples}}
        \left[ y_{t,i} \log \frac{1}{y_{p,i}} + (1-y_{t,i}) \log \frac{1}{1-y_{p,i}} \right] \,.
\end{equation}
See \cref{Sec:Appendix} for a discussion of alternative approaches. The first term of $\ell_{\text{BCE}}$ is identically 0 for events from class 0, while the second term is 0 for events from class 1. Revisiting the example events with $\mathcal{Y} = 3$ and 6 from above, when $y_t$ = 0, the loss is $\ell_\text{BCE}(\{0,0\},\{0.95,0.998\}) \simeq 3+6$. Thus, while the network predicts that both events are signal like, the second event is penalized much more in the loss, because of the network's confidence in the wrong answer. The converse happens when the predicted class is correct. For example, $\ell_\text{BCE}(\{1,1\},\{0.95,0.998\}) \simeq  0.05+0.002$. The BCE loss function gives a comparable very small loss for all reasonably confident, correct predictions, but larger losses for increasingly confident, wrong predictions.

\begin{figure}[tb]
  \begin{center}
    \includegraphics[width= 0.95\columnwidth]{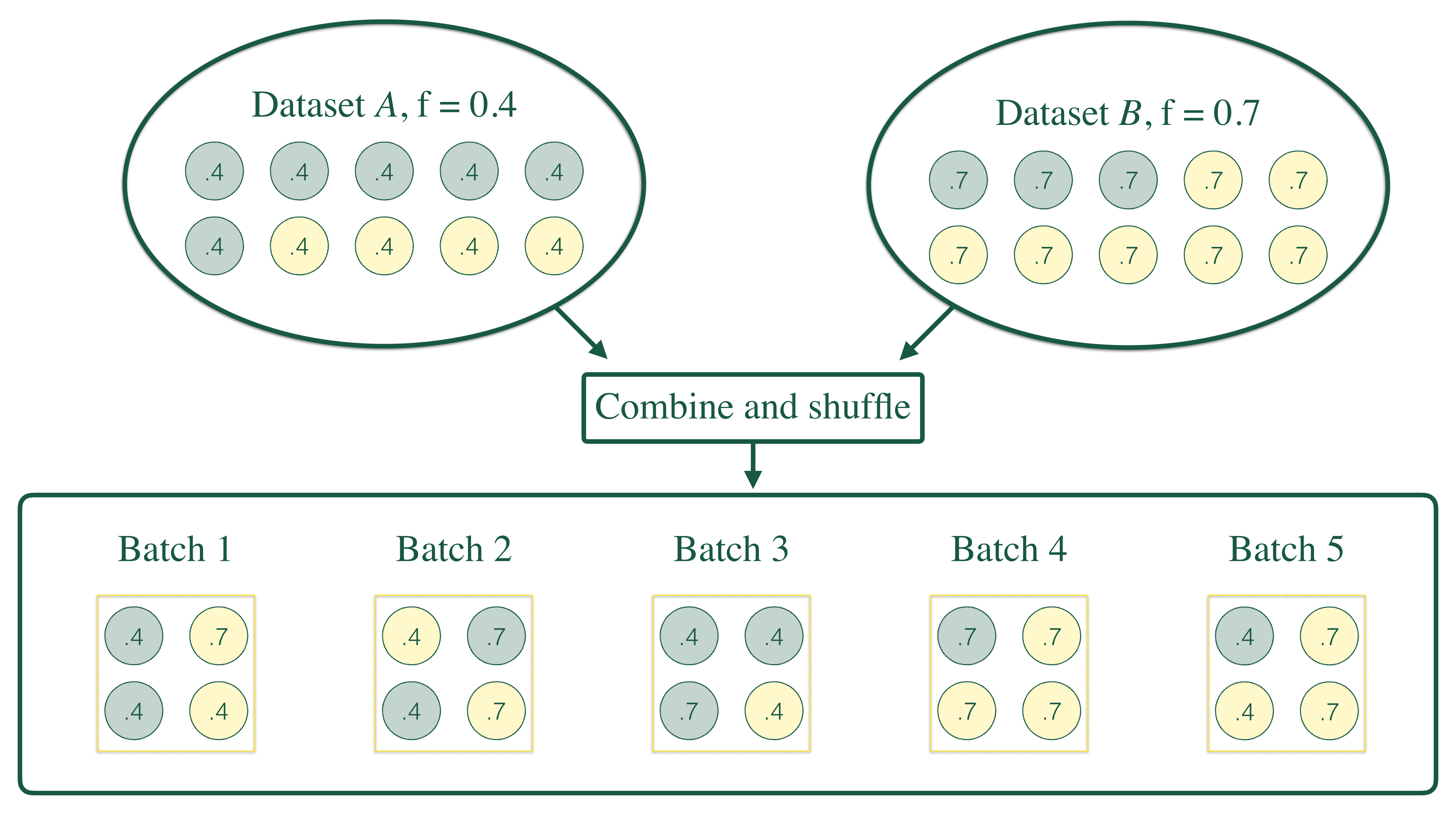}
    \caption{Schematic of the data flow. We are given (at least) two data sets and told to learn two different classes (yellow and green), without knowing the truth value of any one event. Instead, each event is only labeled with the ratio of yellows to  the total that are in its dataset.  The datasets are combined and shuffled into batches, that are used for training.  For illustration in this figure, the batch size is four, and this represents one random permutation.}
  \label{fig:CombineCartoon}
  \end{center}
\end{figure}

For weak supervision, all that is known is the fraction of event classes in the training sample. This fraction contains all the information supplied for the supervision of the training and will be denoted by $f_t$.  However, it will act similarly to $y_t$ --- $f_{t}$ is the same value for all events in a sample, regardless of whether they are from class 0 or 1. For illustration, see \cref{fig:CombineCartoon} where we have two unique datasets, $A$ and $B$, and we want to classify events as green (class 0) or yellow (class 1). The truth level value of each event is not known, only the ratios in the data set, such that each event carries the label $f_t$ from the fraction of yellow circles in its dataset and all of the events from datasets $A$ and $B$ are marked with 0.4 and 0.7, respectively.

Naively, one might expect that since the individual labels of 0 (green) and 1 (yellow) are unknown event by event, \cref{eqn:FullLoss} cannot be used as the loss function --- this motivated the loss function proposed in~\cite{Dery:2017fap}.  However, as demonstrated in~\cite{Metodiev:2017vrx}, one can use any loss function when working with mixed data sets (at least in the infinite statistics limit).\footnote{We are grateful to Eric Metodiev for identifying this aspect of our implementation.}   Using \cref{eqn:FullLoss} for both full and weak supervision allows the most direct comparison of their performance.  In \cref{Sec:Appendix}, we show how the performance of weakly supervised networks depend on the chosen loss function.

Training is done in \emph{batches} (random partitions of the original training set), rather than on the entire combined dataset, which leads to improved computational efficiency.  In \cref{fig:CombineCartoon}, we split the dataset up into 5 batches each with 4 events, denoted by the smaller boxes.  We proceed with training, treating each batch as its own independent dataset.  The weights, $\theta$, of the network are updated using the gradient of the loss function. The amount that the weights change is controlled by the learning algorithm and the learning rate. After updating the weights once for the first batch, we then move to the next batch, where the process is repeated. This continues until the weights have been updated for each of the batches. 

In practice, only running through the training samples once does not allow for enough $\theta$ updates to yield good discrimination, although this does depend on the overall size of the dataset. The process of shuffling the data, splitting it into batches, and training should be repeated many times; this choice, another hyper-parameter, is known as the number of \emph{epochs}, as is the size of the batches. \Cref{tab:Hyperparameters} summarizes all relevant hyper-parameters used for the networks in this work. 

\begin{table}[tb]
  \begin{center}
    \renewcommand{\arraystretch}{1.2}
    \setlength{\tabcolsep}{7pt}
    \setlength{\arrayrulewidth}{.4mm}
    \begin{tabular}{l|cc}
      \hline\hline
      Choice				& Toy Models	& BSM Scenario			\\
      \hline
      Loss function			& BCE		& BCE					\\
      $n_\text{input}$			& 3			& 11						\\
      Hidden Nodes			& 30			& 30						\\
      Activation				& Sigmoid		& Sigmoid					\\
      Initialization				& Normal		& Normal					\\
      Learning algorithm		& \textsc{Adam}		& SGD					\\
      Learning rate			& 0.0015		& 0.01					\\
      Batch size				& 32			& 64						\\
      Epochs				& 100		& 20						\\
      \hline\hline
    \end{tabular}
    \caption{Values of hyper-parameters chosen for the networks used below.  The learning is implemented using \textsc{Keras}~\cite{chollet2015keras} with the \textsc{Adam} optimizer~\cite{DBLP:journals/corr/KingmaB14} for the toy models and stochastic gradient descent (SGD) for the particle physics example.}
  \label{tab:Hyperparameters}
  \end{center}
\end{table}

\Cref{fig:CombineCartoon} has been used as a tool to help visualize the setup of weak supervision, but as such, it is highly idealized. When actually implementing weak supervision on data, the two datasets do not need to be the same size. In addition, if weak supervision is used on real data, the true label of events will be unknown.   In the context of particle physics, the different samples could be obtained by mutually exclusive bins in the transverse momentum or pseudorapidity, for example.\footnote{There is an important assumption which is that the underlying distributions of the features should remain consistent for a given class across the different datasets.  This implies one must be careful when choosing the binning that yields the input data.} In such a situation, theory could predict the ratios, even without having a labeled dataset. However, there are always errors inherent in the predictions, and understanding their impact will occupy much of what follows.
 
Next, we consider a simple toy model and its behavior under the distortion of the input labels. Applications to collider physics are presented later in \cref{sec:BSM}.

\section{Label Insensitivity in Toy Models}
\label{sec:toys}

\begin{figure}[tb!]
  \begin{center}
    \includegraphics[width=\columnwidth]{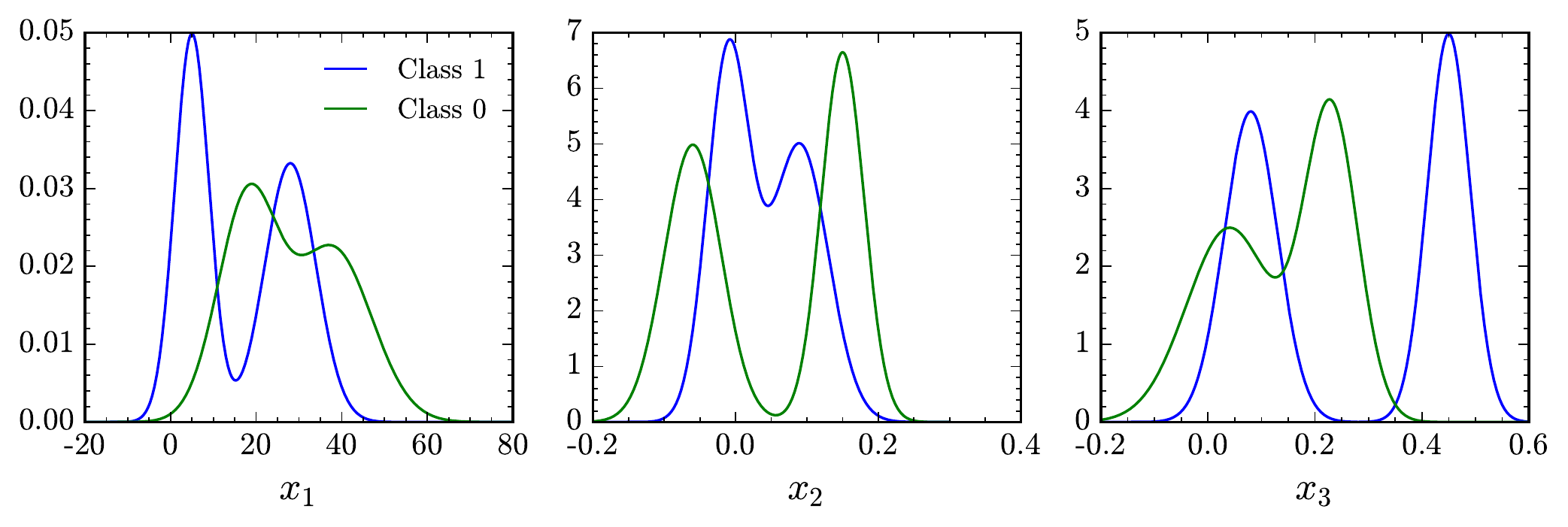}
    \caption{Probability density functions for our toy model. The means and standard deviations of the individual modes are found in \cref{tab:ToyModelDistributions}.}
  \label{fig:ToyModelPDFS}
  \end{center}
\end{figure}

To gain some intuition, we begin by working with simple toy models built from multivariate Gaussians. Distributions for events of class 0 and 1 are produced in three independent variables. To ensure that the impressive performance of the weak learning algorithm are robust to complicated distributions, we make each variable bimodal:
\begin{align}
P(x_i) = \sum_{j=1}^2 \frac{1}{2} \frac{1}{\sigma_i^{(j)} \sqrt{2\,\pi }} \exp\left[ -\frac{1}{2} \left( \frac{x_i-\mu_i^{(j)}}{\sigma_i^{(j)} } \right)^2 \,\right]
\label{eq:P}
\end{align}
These were chosen such that a completely unsupervised clustering algorithm (such as $k$-means \cite{zbMATH03340881,km2,km3}) would not group the events correctly. \Cref{fig:ToyModelPDFS} plots the distributions for each of the features in the toy model utilized for the rest of this section, with their underlying parameters in \cref{tab:ToyModelDistributions}.  The hyper-parameters of the network used in our studies are provided for reference in \cref{tab:Hyperparameters}. Using the same hyper-parameters and training settings for both the fully and weakly supervised networks allows for a more reliable comparison of performance.

The networks are trained on this toy model with two datasets. Both contained \num{200000} samples with fractions of 0.4 and 0.7 of class 1 events. The two training datasets were combined and shuffled (as shown in \cref{fig:CombineCartoon}) with \SI{20}{\percent} of the sample set aside as a validation set. The classifier's performance is then tested on another dataset of the same size with 0.55 of the events in class 1. The \texttt{StandardScaler} of the \texttt{scikit-learn}~\cite{scikit-learn} package is used to normalize and center the data from the training set. These transformations are applied to both the validation and test sets.

To quantify the performance of various classifiers, we construct receiver operator characteristic (ROC) curves. A ROC curve is generated by plotting the true positive rate against the false positive rate, where the cut on the output discriminator $y_p$ varies along the curve.\footnote{Note that constructing a ROC curve requires passing a fully labeled data set through the already-trained network.  We use our test data set for this purpose when constructing all ROC curves in this paper.}  For a given network and set to be classified, the ROC curve is a measure of its performance for binary classification. A perfect classifier would always have a true positive rate of 1.0 with no false positives. Thus, curves pushed to the upper-left corner are better (a diagonal line is then equivalent to a 50--50 guess). One common metric to compare classifiers is to take the integral of these curves, defined as the area under the curve (AUC), with an AUC of 1 being perfect.

\begin{table}[tb]
  \begin{center}
    \renewcommand{\arraystretch}{1.6}
    \setlength{\tabcolsep}{7pt}
    \setlength{\arrayrulewidth}{.4mm}
    \begin{tabular}{c|cccc|cccc}
      \hline\hline
      Feature & $\mu_1^{(1)}$ & $\sigma_1^{(1)}$ & $\mu_1^{(2)}$ & $\sigma_1^{(2)}$
              & $\mu_0^{(1)}$ & $\sigma_0^{(1)}$ & $\mu_0^{(2)}$ & $\sigma_0^{(2)}$ \\
      \hline
      $x_1$ & 26   & 8    & 5     & 4    & 18    & 7    & 38   & 9    \\
      $x_2$ & 0.09 & 0.04 & -0.01 & 0.03 & -0.06 & 0.04 & 0.15 & 0.03 \\
      $x_3$ & 0.45 & 0.04 & 0.08  & 0.05 & 0.23  & 0.05 & 0.04 & 0.08 \\
      \hline\hline
    \end{tabular}
    \caption{Mean and standard deviations of the normal distributions used for the classes 0 and 1. Each feature (variable) is derived from the bimodal distribution in \cref{eq:P} with the events spread evenly between distributions $j = 1,2$.}
  \label{tab:ToyModelDistributions}
  \end{center}
\end{table}

\begin{figure}[b]
  \setlength{\belowcaptionskip}{-18pt} 
  \begin{center}
    \includegraphics[width=0.45\columnwidth]{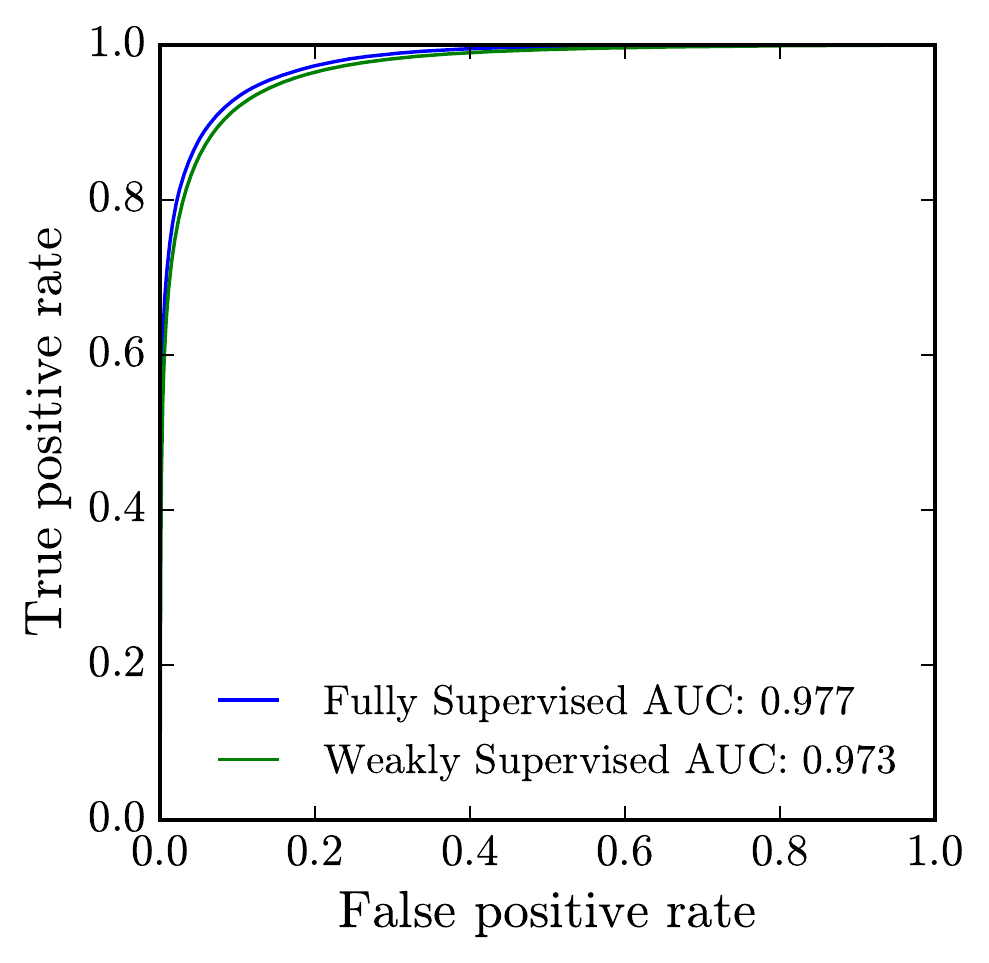}
    \caption{Receiver operator characteristic (ROC) curve for the weakly supervised and fully supervised networks. The performance is very similar for the two networks. }
  \label{fig:ToyRocCurve}
  \end{center}
\end{figure}

\Cref{fig:ToyRocCurve} shows the two resulting ROC curves for the weakly and fully supervised networks on the same training and test data. The AUC for the networks are 0.973 and 0.977, respectively. This serves to validate the claim that weak supervision can yield comparable results to a fully supervised network, despite not knowing event-by-event labels.

To study the sensitivity of the method to accurate label information, we explore how the network performs when there is uncertainty in the fraction labels of the datasets. The same data structure, with dataset fractions 0.4 and 0.7, each with \num{200000} samples, are used throughout. However, for the set with a fraction of 0.7, the actual number fed into the $f_t$ for these samples is varied between 0.0 and 1.0 in steps of 0.1. With weak supervision, changing $f_t$ should result in a change in the prediction. As there are some stochastic steps involved in training the networks (both fully and weakly supervised) we repeat the process 100 times at each value of the label to get a sense of the uncertainty.

\begin{figure}[tb]
  \begin{center}
    \includegraphics[width=\columnwidth]{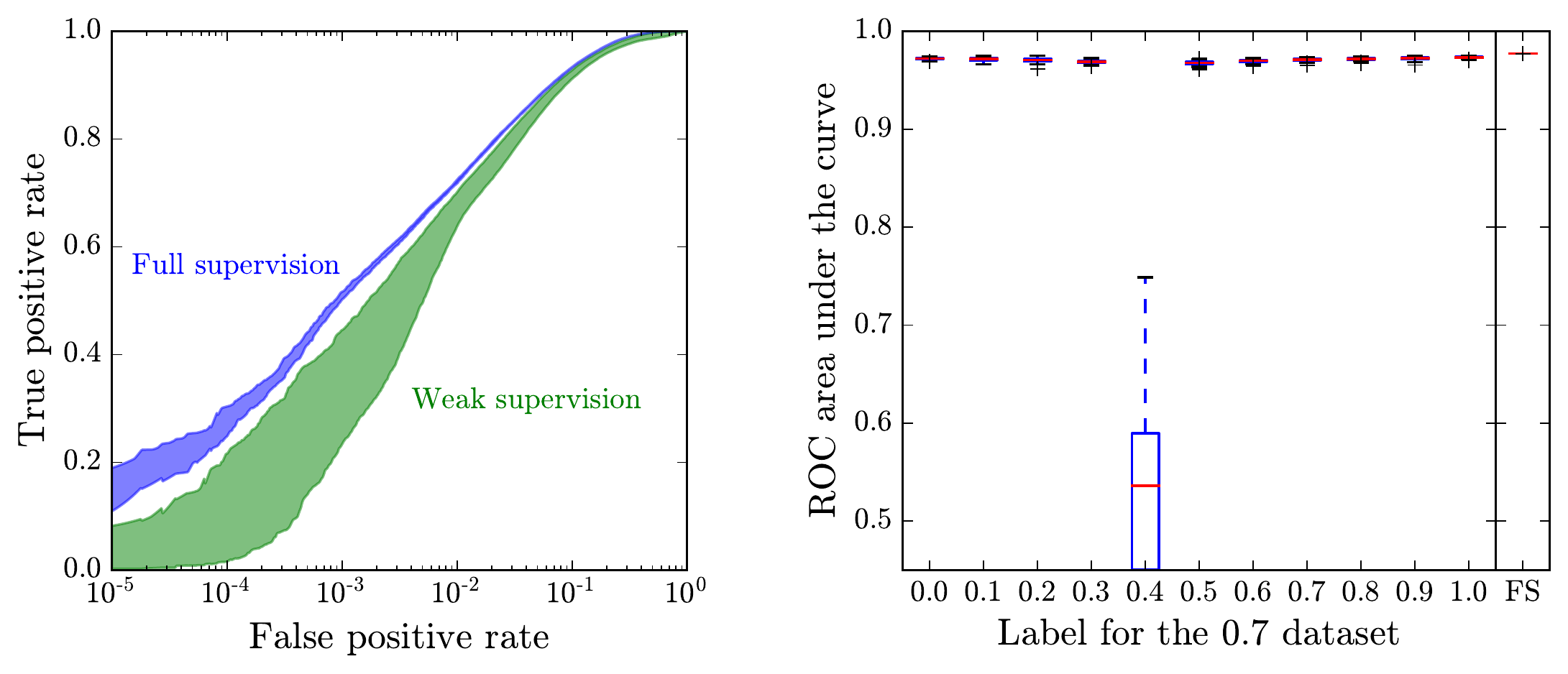}
    \caption{The left panel shows the spread derived from the 100 networks trained to compute the error bars.  We show the result for both fully and weakly supervised networks; for the later, the network is trained using the correct choice of 0.4 and 0.7 for the labels. The area under the curve is very similar for all of them, but the curves differ in their variance and at very low values of the false positive rate, demonstrating the possible spread in network performance trained on the same data. The right panel shows a comparison of the fully supervised network (labeled ``FS'') and the weakly supervised network when one of the datasets labels does not match its true ratio. One data set has a true ratio of 0.4 and is labeled as such. The other dataset has a ratio of 0.7, but the network is trained using the label as specified on the $x$-axis. }
  \label{fig:TwoGausWL}
  \end{center}
\end{figure}

In the left panel of \cref{fig:TwoGausWL}, we have plotted the spread in ROC curves for the different trainings for the weakly supervised networks using the true labels of 0.4 and 0.7, along with the spread of the fully supervised networks. We see that while the integrated area under the curve is very similar for the set, they behave differently when high purity samples are requested. Being a harder classification task, weak supervision displays a greater variance under training than full supervision with the same choice of hyperparameters. Changing the details of the training could lead to a reduction of the difference. The right panel contains  box-and-whisker plots (boxplots) of the AUC for each of the labels using all 100 trainings, along with the 100 instances of the fully supervised network. The boxplots have the red line at the median of the 100 samples, with the boxes extending between the 25\% and 75\% quantile (\SI{50}\% of the samples are in the box). The whiskers extend to the remaining events, unless the point is an outlier, \ie, more than 1.5 out of the quartile range, in which case it is marked by a cross.

Amazingly, even though the mislabeled dataset really has a ratio of 0.7, it is difficult to see any real difference in performance for the mislabeled datasets, with the exception of the 0.4 case.\footnote{For the case where the 0.7 dataset is labeled as 0.4, both datasets are labeled with the same value, so the network defaults to the trivial solution, guessing a value of around 0.4 for every sample in the set. The performance of the network is in fact random.} While they do slightly worse than the fully supervised network, their performance shows remarkable resilience with respect to training set mislabeling.

\begin{figure}[t!]
  \begin{center}
    \includegraphics[width=\columnwidth]{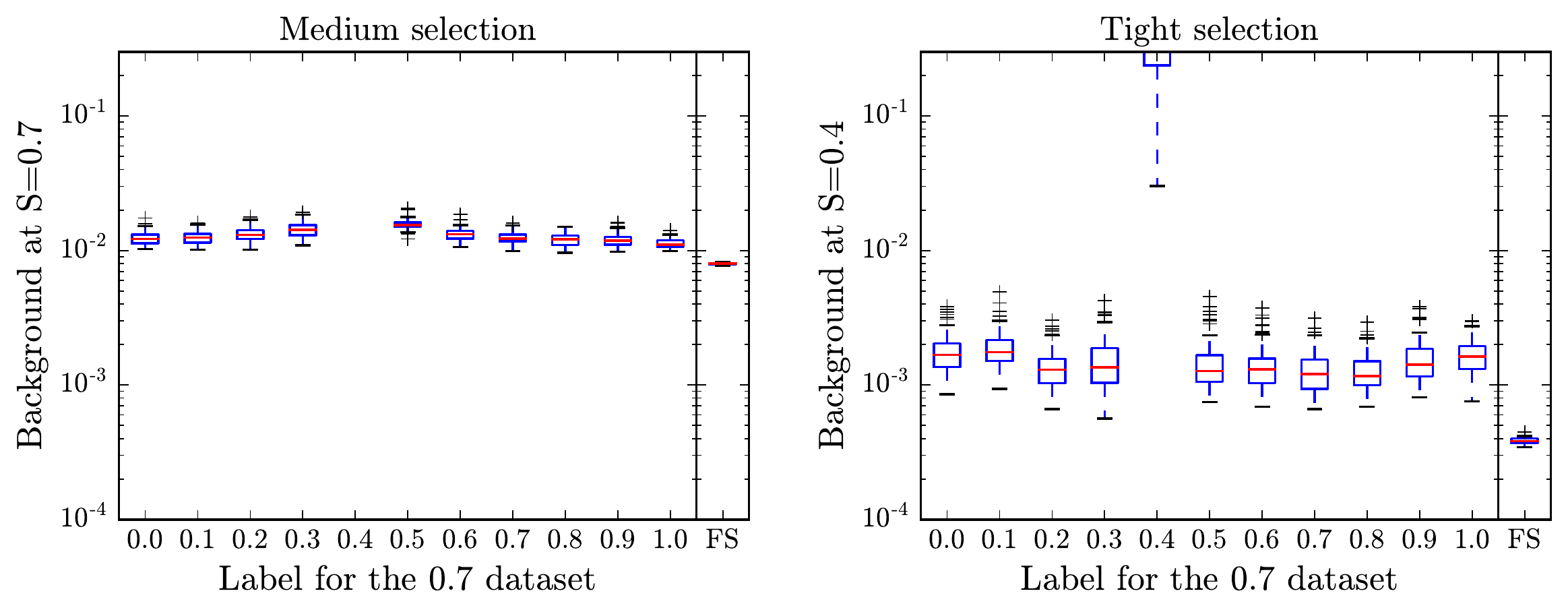}
    \caption{Background efficiencies at fixed values of signal efficiency.}
  \label{fig:WorkingPoints}
  \end{center}
\end{figure}

The AUC gives some sense of how good the classifier is, but actually using the trained network for predictions requires choosing an operating point somewhere along the curve. For instance, tagging of objects at the LHC (such as a $b$-jet or a hadronically decaying $\tau$) often have different working points, where the true positive rate (signal efficiency) have a fixed value. We mimic this, using a signal efficiency of \SI{40}{\percent} as a tight selection, and \SI{70}{\percent} as a medium selection. The best performing network will then allow the least amount of background at these fixed values of signal efficiencies. \Cref{fig:WorkingPoints} shows boxplots of the background efficiency for the different mislabelings at the medium (left) and tight (right) working points. At the medium working point, the median background efficiency for the fully supervised network is around \SI{0.8}{\percent}. The weakly supervised network does slightly worse, achieving background efficiencies around \SIrange[tophrase={--},range-units=single]{1}{1.5}{\percent}.  There is no obvious trend where making the mislabeling worse (far from \SI{0.7}) leads to further degraded performance.

The tight working point, shown in the right panel, has a much stronger background rejection. The background efficiency is around \SI{0.04}\percent ~for the fully supervised networks and \SIrange[tophrase={--},range-units=single]{0.1}{0.2}{\percent} for weak supervision. Weak supervision continues to perform well, regardless of the accuracy of the provided label. In addition, we tested changes to the data set size and choice of activation function on the hidden layer which yielded minimal change in performance. It appears that the fully supervised networks are performing better than the weakly supervised ones, contrary to what would be expected from~\cite{Dery:2017fap, Metodiev:2017vrx}. However, we note that these proofs rely on obtaining the optimal classifier and infinite statistics, which our networks are only an approximation of. In particular, if weakly supervised techniques are trained on real data, as originally envisioned in~\cite{Dery:2017fap}, the data available for training may be smaller than that required to saturate in-principle performance. Likely this would lead to greater variance in training, as observed in \cref{fig:TwoGausWL}, requiring appropriate treatment of systematics and would grow as a concern with input dimensionality. In~\cref{Sec:Appendix}, we show how the choice of loss function affects this approximation. Even without further optimization of the hyper-parameters, we show below that their label insensitivity allows them to outperform fully supervised networks in the presence of mismodeling.

\begin{figure}[tb]
  \begin{center}
    \includegraphics[width=\columnwidth]{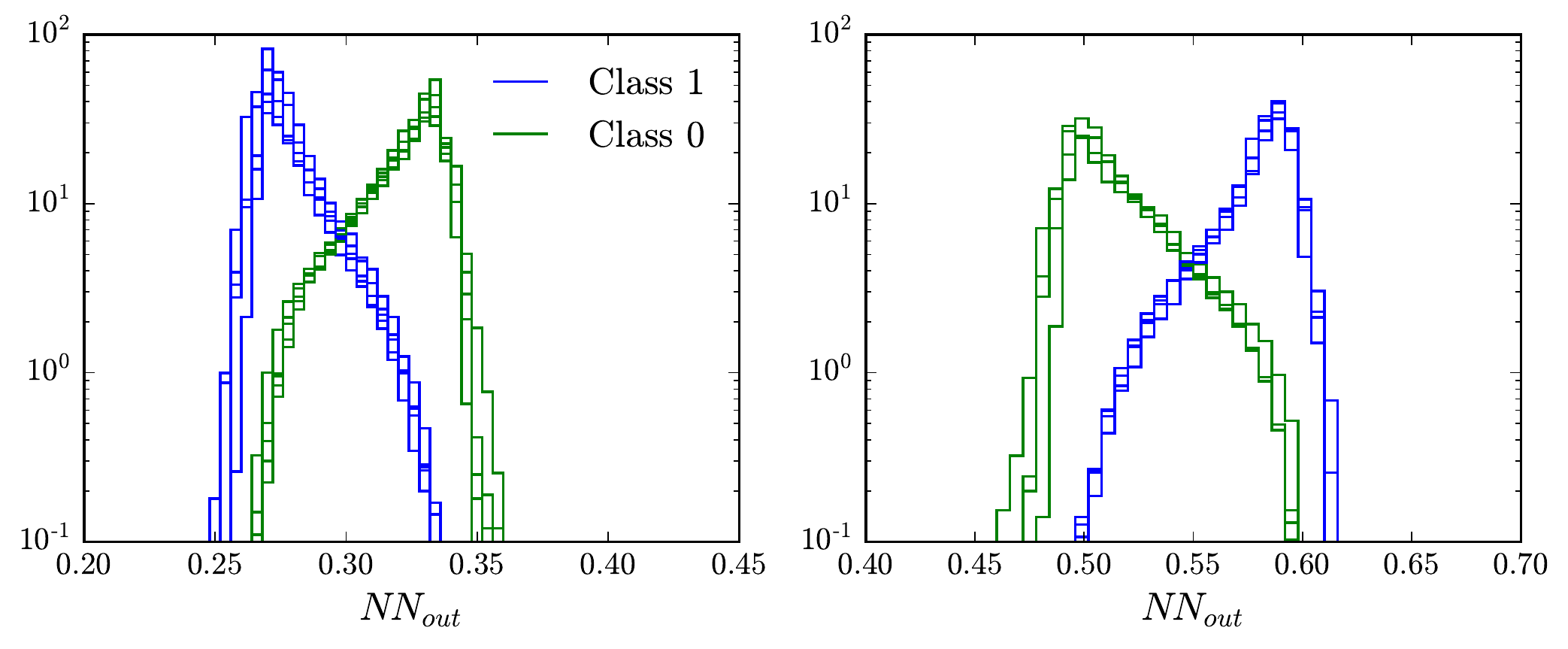}
    \caption{Outputs of the weakly supervised neural network. The blue lines show distribution of signal for 10 separate trainings, while the green lines show the background. The left and right panels represent when the data set which is has a truth value ratio of 0.7 is labeled as either 0.2 or 0.7, respectively. As pointed out in~\cite{Dery:2017fap}, the possibility of class-flipping exists for weakly supervised networks. To get the signal events in the left panel, the cut on the network output is an upper bound, while in the right panel it is a lower bound.}
  \label{fig:ToyModelOutputs}
  \end{center}
\end{figure}

A suggestive aspect of the classifier resilience can be seen in \cref{fig:ToyModelOutputs}, which shows 10 overlaid histograms of the neural network outputs for the signal and background test samples. The right panel that sample is labeled with the true fraction, and the left panel shows distributions for when the data set which has a true fraction of 0.7 is mislabeled as 0.2. We see that the network changes the preferred location for the signal samples, but that the classifier distributions of events by class look like rescaled versions of each other. Thus, while the predicted value for any given sample may change, the network is still able to tell a difference between signal and background. 

The results of this section have focused on when only one of the fraction labels is mislabeled. However, it is realistic to think that an uncertainty in the fraction could extend to both datasets. We have checked multiple ways of mislabeling both datasets including both fractions shifting up or down and the fractions shifting towards and away from each other. In each instance, we find that the results presented here hold, and there is no significant change in the performance of the network. The next section examines the conditions such that weakly supervised networks can remain so robust to input fraction errors.

\section{Analytic Condition for Label Insensitivity}
\label{sec:Analytics}

The behavior of the classifier with respect to a mislabeling of the fractional labels seems  surprising. After all, it is as though we have removed essentially all supervision from our learning algorithm without any significant degradation of performance. This is true, although only to a point. This insensitivity to inputs can be traced to the fact that the optimal classifier is not, in fact, unique, but rather parametrized by a family of functions related by all possible monotonic mappings. The putative optimal classifier that an algorithm reconstructs can consequently suffer from quite large distortions without degrading its overall performance. However, the mislabeling of the training data does extract a price, as the particular cut associated with a given working point will end up misidentified. These claims are made more precise below, along with giving a relatively simple breakdown criterion at which performance will begin to worsen.

First, we consider the behavior of an optimal classifier given correct inputs and infinite training statistics. For given inputs and dataset to be classified, the optimal classifier can be constructed from the best approximation of an event's probability being in a given event class at every point in feature space. The ratio of these probabilities can then act as a discriminating variable that can be cut on. We frame the argument in the classification approach of~\cite{Dery:2017fap}, in contrast to the method adopted in the numerical implementation here and in~\cite{Metodiev:2017vrx}. Because, in the former case, classifier training and cut selection is accomplished in one step, the generic universal aspects of output behavior is more easily understood. The latter divide the task of training and associating rejection rates with specific cuts on classifier output into separate steps. The nature of the resulting errors, when present, then have dependence on the precise methodology adopted. The statements about breakdown of optimal performance that we derive below are applicable to both cases, although the precise behavior after breakdown will depend on the class of loss function chosen.

Let us consider a discretization of our input features. Every training set then corresponds to an $n$-dimensional histogram, $n$ being the number of features. Referring to each bin by the collective coordinate $i$, the distributions for each event class can then be found by inverting the known distributions on a bin-by-bin basis.\footnote{With more than two training samples, the system is over-constrained. Analytically this could be dealt with by solving the resulting equations with respect to some test statistic, \eg, least-squares. In a machine learning context, the training procedure packages together the freedom associated with this choice with that of approximating the distributions over feature space into one step.} Here, $h_{A,i}, h_{B,i}$ denote the number of events in bin $i$ of training sets $A, B$, and $f_{A,B}$ the total fraction of event class 1 in each set. 
\begin{equation}
\label{eqn:DistInv}
  \begin{matrix}
    h_{A,i} = f_A h_{1,i} + (1 - f_A) h_{0,i} \\[5pt]
    h_{B,i} = f_B h_{1,i} + (1 - f_B) h_{0,i}
  \end{matrix}
    \quad \Longrightarrow \quad
  \begin{matrix}
    h_{0,i} = \frac{f_A h_{B,i} - f_B h_{A,i}}{f_A - f_B} \\[10pt]
    h_{1,i} = \frac{(1 - f_B) h_{A,i} - (1 - f_A) h_{B,i}}{f_A - f_B}
  \end{matrix}
\end{equation}
In the limit of infinite statistics, this inversion will accurately compute the event class distributions in every bin. An optimal classification could then be achieved given by cutting on 
\begin{equation}
  \bar{z}_i = \frac{h_{1,i}}{h_{0,i} + h_{1,i}}\,.
\end{equation}
Crucially for our results, this choice is not unique. 

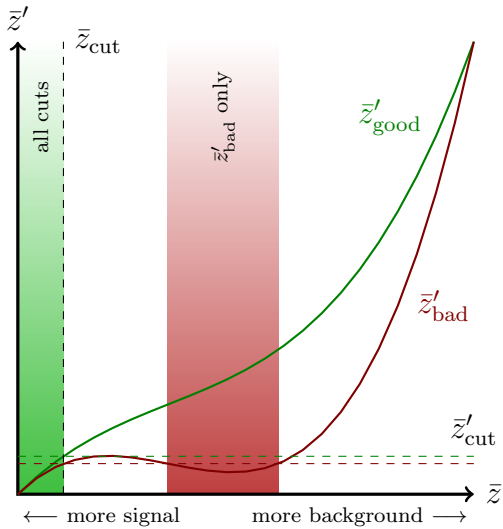
\begin{figure}[tb]
  \begin{center}
    \begin{tikzpicture}[scale=2]
	  \shade[top color=white, bottom color = green!50!gray]
	    (0,0) -- ++(0.3,0) -- ++(0,3) -- ++(-0.3,0) -- cycle;
		
	  \shade[top color=white, bottom color = red!50!gray]
		(0.986,0) -- ++(0.728,0) -- ++(0,3) -- ++(-0.728,0) -- cycle;
		
	  \draw[<->,very thick] (0,3) node (yaxis) [above] {$\bar{z}'$}
		|- (3,0) node (xaxis) [right] {$\bar{z}$};
	  \node at (0.55,-0.15) {\scriptsize $\longleftarrow$ more signal};
	  \node at (2.25,-0.15) {\scriptsize more background $\longrightarrow$};
		
	  \draw[color=green!50!black, thick]
		plot[domain=0:3] ({\x}, {\x - 0.2*\x^2*(3-\x)});
	  \node at (2.45,2.5) {\color{green!50!black}{$\bar{z}'_\text{good}$}};
		
	  \draw[color=red!50!black, thick]
		plot[domain=0:3] ({\x}, {\x - 0.4*\x^2*(3-\x)});
	  \node at (2.8,1.25) {\color{red!50!black}{$\bar{z}'_\text{bad}$}};
		
	  \draw[dashed] (0.3,0) -- (0.3,3) node [right] {$\bar{z}_\text{cut}$};
	  \draw[color=green!50!black, dashed] (0,0.2514) -- (3,0.2514)
		node [above] {\color{black}{$\bar{z}'_\text{cut}$}};
	  \draw[color=red!50!black, dashed] (0,0.2028) -- (3,0.2028);
		
	  \node at (0.15,2.5) {\scriptsize \rotatebox{90}{all cuts}};
	  \node at (1.35,2.5) {\scriptsize \rotatebox{90}{$\bar{z}'_\text{bad}$ only}};
	\end{tikzpicture}
	\caption{The effect of a cut on a deformed classifier $\bar{z}'_i$ in terms of cuts on an optimal classifier $\bar{z}_i$. As long as the mapping between the two is monotonic, a single cut on $\bar{z}'_i$ will always correspond to some cut on the optimal classifier. As soon as monotonicity breaks down, some cuts on $\bar{z}'_i$ will be contaminated by less signal-pure regions of feature space.}
	\label{fig:AnalyticSketch}
  \end{center}
\end{figure}

Why this should be so is illustrated in \cref{fig:AnalyticSketch}. Since $\bar{z}_i$ is an optimal classifier by construction, a single cut on $\bar{z}_i$ will correspond to the best possible purity that can be achieved for a given signal acceptance rate. As long as the mapping from $\bar{z}_i$ to $\bar{z}'_i$ is monotonically in- or decreasing over the range of the original classifier, in this case $\bar{z}_i \in [0,1]$, a single cut on $\bar{z}'_i$ will correspond to some cut on the optimal classifier. As soon as the monotonicity breaks down, there will be some choice of cut on $\bar{z}'_i$ for which less signal-pure regions of feature space are included without optimally increasing signal acceptance. If we can identify the point at which monotonicity breaks down as a function of input mislabeling, we can predict at what point performance will degrade.

The explicit form of the optimal classifier allows us to see why mislabelling the training sets has so little effect. From \cref{eqn:DistInv}, we write
\begin{equation}
  \bar{z}_i
    = \frac{h_{1,i}}{h_{0,i} + h_{1,i}}
    = \frac{(1 - f_A) h_{B,i} - (1 - f_B) h_{A,i}}{(1 - 2f_A) h_{B,i} - (1 - 2f_B) h_{A,i}}
    = \frac{1-f_B}{1-2f_B} \frac{\frac{1 - f_A}{1 - f_B} - r_i}{\frac{1 - 2f_A}{1 - 2f_B} - r_i},
\end{equation}
where we have defined $r_i = h_{A,i}/h_{B,i}$. As one would expect, the classifier checks if $r_i$ is close to the ratio one would expect for a region dominated by one class and returns a value close to that class. The mislabeling of \cref{sec:toys} could then be captured by a fractional shift $f_A \to f_A + \delta$, leading to the classifier being reconstructed as
\begin{multline}
\label{eq:zShift}
  \bar{z}'_i = \frac{1-f_B}{1-2f_B} \frac{\frac{1 - f_A - \delta}{1 - f_B} - r_i}
                                         {\frac{1 - 2f_A - 2\delta}{1 - 2f_B} - r_i}
             = \bar{z}_i + \delta \left(
                 \frac{\frac{1-f_B}{1-2f_B} - \frac{\bar{z}_i}{1-2f_B} + 2(\bar{z}_i^2-\bar{z}_i)}
                      {\frac{f_A-f_B}{1-2f_B} + 2\delta(\frac{1-f_B}{1-2f_B} - \bar{z}_i)}\right) \\[10pt]
             = \bar{z}_i + \delta \left( \frac{(1-f_B) - \bar{z}_i + 2(1-2f_B)(\bar{z}_i^2 - \bar{z}_i)}
                                              {f_A - f_B} \right) + \ord(\delta^2).
\end{multline}
One thing that is immediately clear from the expression above is that sensitivity to mislabeling increases as $f_A$ and $f_B$ approach each other. If the mapping from $\bar{z}_i$ to $\bar{z}'_i$ is given by a monotonic function for positive $\bar{z}_i$, then by the discussion above, they actually correspond to the same optimal classifier, and no loss of performance will occur. Since $\bar{z}_i$ can in principle take on any value between 0 and 1, to determine monotonicity we are justified in treating it as a continuous parameter, so that
\begin{multline}
  \frac{\text{d}\bar{z}'_i}{\text{d}\bar{z}_i}
     = 1 + \delta
           \left( \frac{\frac{f_A-f_B}{1-2f_B}\left( 4\bar{z}_i-2 - \frac{1}{1-2f_B} \right)
           	            + 2\delta \left( \frac{1-f_B}{1-2f_B}(4\bar{z}_i-1 - \frac{1}{1-2f_B}) \right)
           	            - 2\bar{z}_i^2} 
                       {\left( \frac{f_A-f_B}{1-2f_B}
                       	       + 2\delta(\frac{1-f_B}{1-2f_B} - \bar{z}_i) \right)^2} \right) \\[10pt]
     = 1 + \delta \left( \frac{(1-2f_B)(4\bar{z}_i - 2) - 1}{f_A - f_B} \right) + \ord(\delta^2)
\end{multline}
Assuming that $f_A > f_B$, the minimum value of the second term over the physical range of $\bar{z}_i$ is approximately given by $ -\delta (3 - 2\min(f_B, 1-f_B))/(f_A - f_B)$. For sufficiently small $\delta$, the only mappings suffering no distortions will be ones that are monotonically increasing, in which case the requirement would be that the second term remains greater than $-1$, \emph{i.e.},
\begin{equation}
  \delta \lesssim \frac{f_A - f_B}{3 - 2\min(f_B, 1-f_B)}.
\end{equation}
A more careful treatment would consider solutions with monotonically decreasing mappings allowed, but in all cases, as stressed with \cref{fig:AnalyticSketch} above, the mapping from $\bar{z} \rightarrow \bar{z}'$ must be monotonic if it is to stay within the family of optimal classifiers.

What is happening here is that in the limit of infinite statistics, the weakly supervised classification problem really has an analytically optimal solution. But by distorting the inputs, the event class probabilities at each point in feature space seen by the learning algorithm are incorrect. However, this only causes improper separation of signal and background if feature space points with different event class probabilities are mapped to the same incorrect probabilities. As captured in \cref{eq:zShift}, this requires a combination of training sets that are similar in their composition together with a sufficiently large mislabelling. As a result, for most distortions the actual contours of the classifier output swept out in feature space do not change, only the cut values associated with them. The issue of the wrong working point being used when a particular signal strength is required still remains, but it will correspond to a different optimal working point rather than a suboptimal classifier.

\section{Label Insensitivity for LHC Physics}
\label{sec:BSM}

Part of the motivation provided in~\cite{Dery:2017fap} for the use of weakly supervised networks is that they can, in principle, be trained on real data, and thus be robust against modeling errors that could lead to changes in the fractions. This strategy applies well to tagging SM objects, demonstrated therein, using weakly supervised networks to distinguish between quark and gluon jets. This motivation may not apply to searches for BSM physics, where there is no way to guarantee that new physics is in the real data, let alone that we know the ratio.\footnote{It is interesting to consider if this could be used to help distinguish between BSM scenarios when a large excess is discovered.  We leave studying this application to future work.} Training on backgrounds in a data-driven way could still be accomplished by injecting signals into real data. Moreover, as explored in the following, the use of weakly supervised networks could provide a safeguard both against mismodeling and signal--background contamination in BSM searches.

\subsection{Gluino versus \emph{Z} + jets}
\label{sec:BSM_initial}
The particular BSM scenario we choose to study is the canonical gluino--neutralino simplified model. We set the mass of the gluino to be \SI{2}{\TeV}, which is near the current limits~\cite{ATLAS-CONF-2017-022, CMS:2017gzh}. The gluinos are pair produced, and then decay to the neutralino (which we take to be massless) and a pair of quarks. This yields a signature of many jets and missing energy. One of the dominant backgrounds for these types of searches is $Z + \text{jets}$, with $Z$ decaying to neutrinos. As this is a proof of concept, rather than a dedicated search, we only use this background, with one hard jet in the initial interaction.

\textsc{Madgraph5}~\cite{Alwall:2014hca} is used to generate the signal and background Monte Carlo samples. The events are showered and hadronized using \textsc{Pythia6.4}~\cite{Sjostrand:2006za}. Detector simulation is done with \textsc{Delphes3}~\cite{deFavereau:2013fsa} using the default detector card, modified to use anti-$k_t$ jets~\cite{Cacciari:2011ma} with a radius of $R = 0.4$, as calculated with \textsc{FastJet}~\cite{Cacciari:2008gp}. We initially generate \num{500000} events for the signal and \num{1000000} for the background. The background samples have an extra, generator-level cut such that the jet has  $p_T > \SI{150}{\GeV}$ and the minimum missing energy is \SI{150}{\GeV}. At the detector level, we impose an additional cut so that both the missing energy and the transverse momentum of the hardest jet is greater than \SI{200}{\GeV}. For any jet to be considered, it must have $|\eta| < 2.5$, and we record the $p_T$ of up to 10 jets, as long as they have $p_T> \SI{40}{\GeV}$. Additionally, any events that contain an isolated lepton are vetoed. 

Of the $10^6$ Monte Carlo events for the background, \num{264303} passed the cuts. Meanwhile, \num{473359}  out of the \num{500000} signal events pass the initial cuts.
Out of all the events passing the cuts, \SI{10}{\percent} are set aside to use as a test set. The remaining are split \SI{80}{\percent} for training and \SI{20}{\percent} for validation. 

We first start with fully supervised training with the goal of distinguishing between the background and the gluino pair production. The data from the training set is centered and normalized using the \texttt{StandardScaler} function in the  \texttt{scikit-learn} package~\cite{scikit-learn}, which is likewise applied to the validation and test data.  A fully supervised network is trained within the \textsc{Keras} framework~\cite{chollet2015keras} using one hidden layer with 30 neurons, where each layer is initialized with the normal distribution and uses a sigmoid activation. We use 11 inputs for the network, the missing transverse energy along with the transverse momentum of the first 10 jets passing the selection cuts. If there are less than 10 jets, the corresponding input is 0.  The minimization is done using SGD with a learning rate of 0.01.

Next, we do the analysis on the same gluino and background events but use weak supervision. To do this, we split the data into two different training sets. The first set contains \SI{80}{\percent} of the background and \SI{40}{\percent} of the signal events. The second set has the remaining \SI{20}{\percent} of the background and \SI{60}{\percent} of the signal. The validation sets are split in the same fashion. This leads to the ratios of the two data sets being 0.47 and 0.83, respectively. The weakly supervised network is then trained in an identical fashion as the fully supervised.

\begin{table}[t]
  \begin{center}
    \renewcommand{\arraystretch}{1.6}
    \setlength{\tabcolsep}{7pt}
    \setlength{\arrayrulewidth}{.4mm}
    \begin{tabular}{clc}
      \hline\hline
      Network	& \multicolumn{1}{c}{AUC}	& Signal efficiency		\\
      \hline
      Full		& \num{0.99992393(31)}		& \num{0.999373(17)}	\\
      Weak		& \num{0.9998978(35)}		& \num{0.999286(30)}	\\
      \hline\hline
    \end{tabular}
    \caption{Metrics for training networks to distinguish gluino pair production with decays to 1st generation quarks from the dominant $Z + \text{jet}$ background. The signal efficiency is given for a background acceptance of 0.01.}
  \label{tab:BSMMetrics}
  \end{center}
\end{table}

Both networks are again excellent classifiers, exemplified by the area under the ROC curves. However, as argued earlier, this is not the best measure for the performance of a classifier in the context of a particle physics experiment. Instead, here we are imagining using the networks to reject backgrounds in some particular analysis. In this case, we want to choose an operating point where a small (fixed) amount of background would be accepted, with the signal efficiency maximized. The test set (as described above) has around \num{20000} background events total. We set the cut such that \SI{1}{\percent} of background event pass and train 10 separate times. \cref{tab:BSMMetrics} shows the resulting signal efficiencies for this background rejection.

\begin{figure}[t]
  \begin{center}
  \includegraphics[width= \columnwidth]{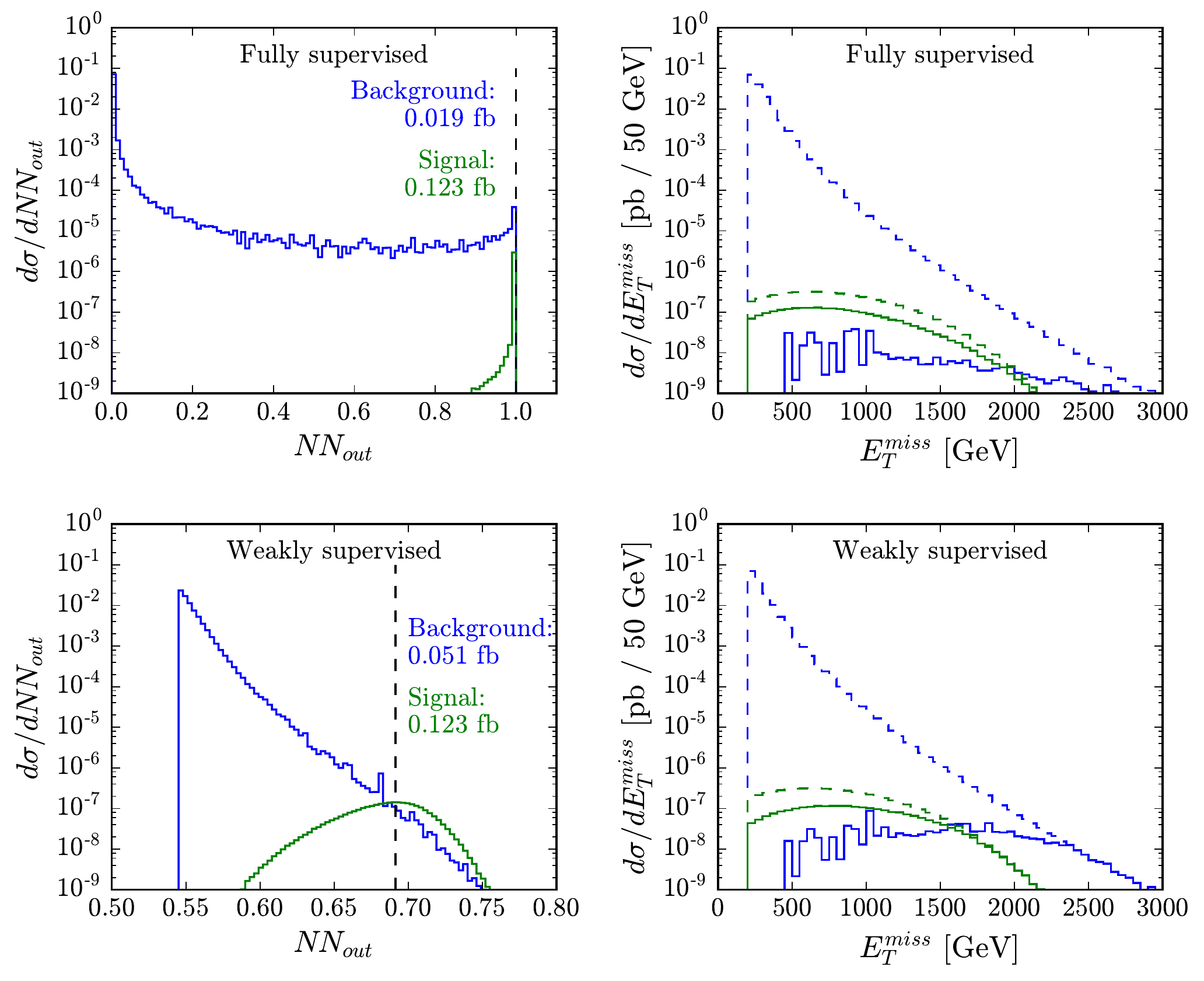}
  \caption{The left panels show the differential cross section as a function of the neural network outputs for the fully and weakly supervised networks on the top and bottom, respectively. The dashed vertical lines show where the cut is placed, while the labels mark the signal and background cross sections passing the cut. The cut was chosen such that both networks yield the same number of signal events.  Note that the top left plot confusingly never shows $S/B > 1$ due to a finite bin size artifact. The right panels show the effect of the cut on the missing energy spectrum. The dashed (solid) lines show the distributions before (after) the cut is made. Both of the neural networks are able to reject the dominant background at low values of missing energy, while keeping most of the similar signal events.}
  \label{fig:BSMMetrics}
  \end{center}
\end{figure}

The networks trained above were based on \num{e6} Monte Carlo events. However, the cross sections for these processes are so large that the effective luminosity for these events is not large enough to span the full relevant phase space. For example, the background in the original dataset has a maximum missing energy of about \SI{1}{\TeV}. However, if we were to simulate the same integrated luminosity for the signal and background, the background could have missing energy as large as \SI{3}{\TeV}.

To account for this, we further generated another sample of backgrounds using generator level bins of missing energy to get non-trivial statistics for rare events. The new events contain much larger missing energy and jet momenta for the background events than were realized in any of the training examples. \Cref{fig:BSMMetrics} shows the resulting predictions for these events from the two networks, weighted by their cross section. The networks again behave similarly, and can effectively separate the signal from the background.  This is very impressive as the background events are probing different regions of phase space than were spanned in the training set.

As a final probe, we next make even more stringent cuts on the neural network outputs (while maintaining that the signal cross section making it through each network is the same). Background cross sections of \SIlist{0.019;0.051}{\fb} are achieved for the fully and weakly supervised networks, with \SI{0.123}{\fb} of signal. The right panels of \cref{fig:BSMMetrics} show the differential cross sections as a function of the missing energy. The dashed lines represent the distributions before making the cuts, while the solid lines are after the cut is made. We see that both networks gain their separating power by cutting out background with relatively small amounts of missing energy.

\subsection{Mismodeling}

Armed with these concrete comparisons between weakly and fully supervised networks, we will explore an example of mismodeling that would lead to a change in the fraction labels provided at the training step. In particular, we will see that for the class of mismodeling effects we study here, the performance of a fully supervised network degrades, while the weakly supervised networks remain robust.

\begin{figure}[b]
  \setlength{\belowcaptionskip}{-18pt}
  \begin{center}
    \includegraphics[width=0.48\columnwidth]{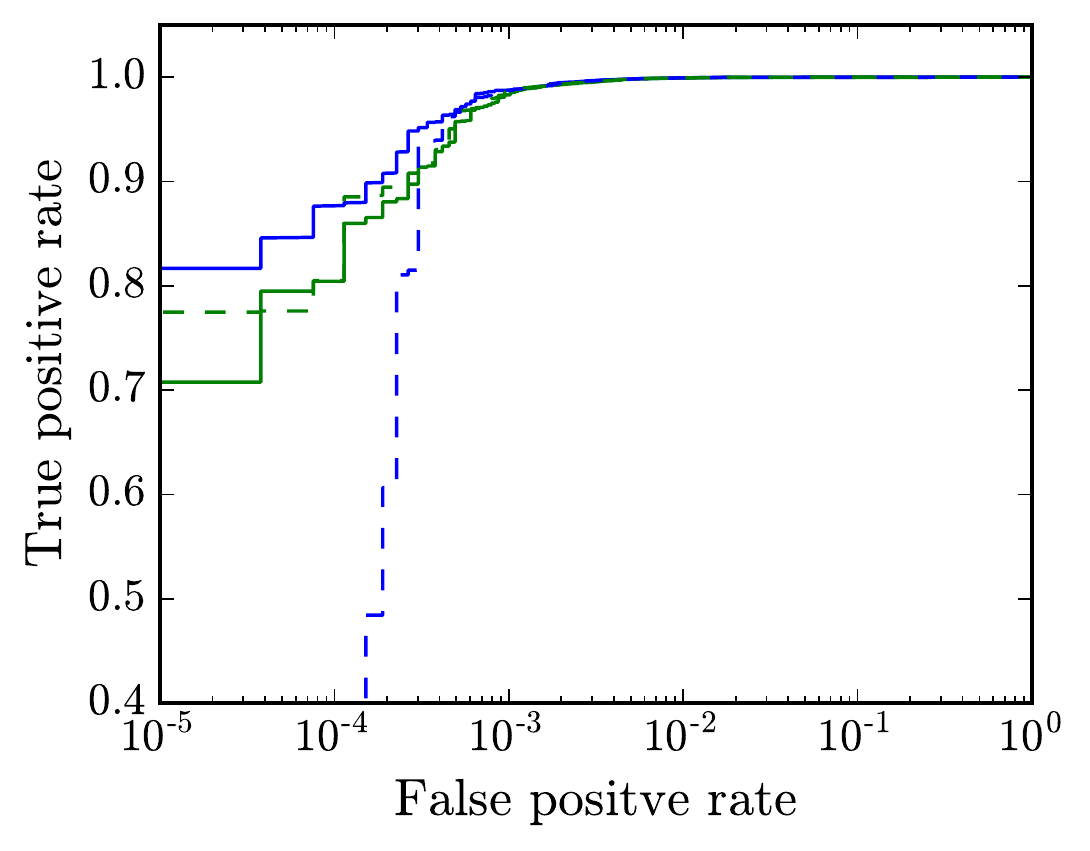} 
	\includegraphics[width=0.48\columnwidth]{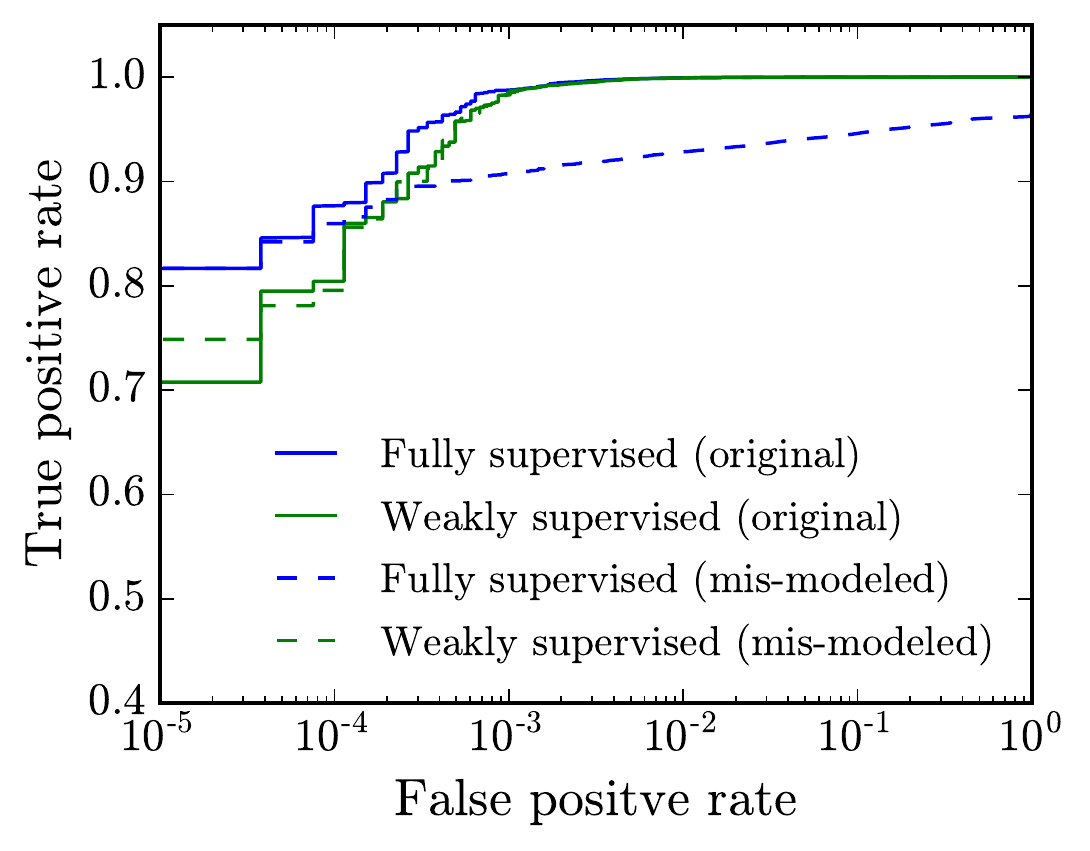}
	\caption{ROC curves showing the response of the network to mismodeled data. The mismodeling in the left panel shows the results of taking a random \SI{15}{\percent} of the signal events and labeling them as background before training. The right panel demonstrates what happens under a phase space swap, where we mislabel the \SI{10}{\percent} most signal-like background event and the \SI{15}{\percent} most background-like signal events. The fully supervised network trained on the mislabeled data performs much worse at small false positive rates than when the data is not mislabeled. The weakly supervised network does not rely on individual event labels, and therefore has essentially no change in performance.}
		\label{fig:Mismodeled}
	\end{center}
\end{figure}

In order to mock up the effects of this mismodeling, we take the original set of training and validation events and use the fully supervised network to classify them. Two tests are then performed and their results are presented in \cref{fig:Mismodeled}. In the first, \SI{15}{\percent} of the signal events are chosen at random and artificially mislabeled as background (left panel). In the second, we perform a \emph{phase space swap}.  Specifically, we change the labels between the  most-signal like \SI{10}{\percent} of the background events and the most background-like \SI{15}{\percent} of the signal events (right panel). These two tests alter the fractions used for the weakly supervised classification. They simultaneously change the underlying missing energy and jet momentum distributions for the training samples in different ways. The events are then split into subgroups as was done above to make samples with different ratios, where the ratio is now calculated based on the updated mismodeled labels. A new weakly supervised network is trained on these misclassified events as well by reporting a new incorrect signal fraction corresponding to the number of signal events flipped for the fully supervised case. \cref{table:BSMMismodel} shows the values of the fractions used in the trainings, as well as the fractions that are actually present in the datasets.

\begin{table}[tb]
	\begin{center}
	   \renewcommand{\arraystretch}{1.6}
    \setlength{\tabcolsep}{7pt}
    \setlength{\arrayrulewidth}{.4mm}
		\begin{tabular}{c | c c c}
			\hline
			\hline
			&  \multicolumn{3}{c}{$f_{t}$ label} \\
			\cline{2-4}
			Dataset & Original & Random 15$\%$& Phase space swap \\
			\hline
			$A$ & 0.472 & 0.374 (0.585) & 0.416 (0.593) \\
			$B$ & 0.843 & 0.782 (0.769) & 0.810 (0.747)\\
			\hline
			\hline
		\end{tabular}
		\caption{Labels for the fractions of the two datasets used when the data is mismodeled. The numbers in parenthesis show the true fraction contained in the dataset.}
		\label{table:BSMMismodel}
	\end{center}
\end{table}

The ROC curves for the new networks when tested on the true (not altered) labels are shown in \cref{fig:Mismodeled}. The new fully supervised network shows a distinct drop in performance when trained on the mismodeled data. When the data is mismodeled, the area under the ROC curve is effectively unchanged for the \SI{15}{\percent} mislabeling test, and is \SI{4.1}{\percent} smaller for the phase space swap test.  Even though the AUC does not change in the left panel, the true positive rate is impacted. However, the weakly supervised network shows little change in performance because it does not rely on individual event labels.

From the point of view of the weakly supervised network such mismodelling looks like fraction mislabelling up to small corrections, and so the same robust behavior is observed. (This is in addition to any explicit fraction mislabelling that would be present in training samples due to, \eg, uncertainties in absolute cross section.) Approaches to mitigation of systematic uncertainties in fully supervised cases exist, but they require additional steps to explicitly take such systematics into account as well as some model for the uncertainties. Weakly supervised networks can achieve better results in the presence of such systematic errors than fully supervised networks when no explicit systematics mitigation technique is used. This result, along with the robustness shown in \cref{fig:TwoGausWL}, motivates further exploration of weakly supervised networks at the LHC.

\section{Discussion and Future Directions}
\label{sec:disc}
In this paper, we have studied the robustness of weakly supervised neural networks.  We confirmed that weakly supervised networks are essentially as performant as fully supervised networks. We further demonstrated that weakly supervised networks are robust to a class of systematic mismodeling effects, and provided an analytic argument to explain this unexpected feature.  To our knowledge, our work is the first application of these kinds of networks to beyond the Standard Model collider search scenarios.  

There are many future directions left to explore.  For example, we provided an oversimplified background sample by only simulating the dominant process, but in practice one should train using a more realistic background involving all relevant Standard Model processes.  We did some very simple explorations of the systematics of mismodeling, but there is clearly much left to explore.
For a study comparing different loss functions, see \cref{Sec:Appendix} and~\cite{Metodiev:2017vrx}.  Additionally, one could tune hyper-parameters, use weight decay \cite{Krogh:1991:SWD:2986916.2987033}, and so on; a full optimization study should also be performed.

It would also be interesting to explore applications to subtle signatures of new physics.  Data from the LHC is used to tune Monte Carlo event generators, which are subsequently used to simulate backgrounds for the LHC.  It is therefore conceivable that a marginal signal of new physics from previously recorded data could be absorbed by this procedure.  This would imply that Standard Model distributions would be distorted by the inclusion of some BSM signal.   Examples where this could occur are models with quirks or hidden valleys, where the BSM physics might end up lurking in very soft hadronic data or a complex hadronic final state~\cite{Strassler:2006im, Kang:2008ea, Harnik:2008ax, Knapen:2016hky}.  It is plausible that weak supervision could be used to tease out these difficult signals without an over-reliance on simulations, although additional work is required to develop an explicit search strategy in this case.

We will conclude with one final observation.  Although this paper was largely devoted to pitting them against each other, it turns out that fully and weakly supervised networks are complementary.   While the performance of both networks has been shown to be very similar, on an event-by-event basis, the two networks are relying on different events for maximal discrimination power.  This implies that even better classification is possible using the combined output from both types of networks.

\begin{figure}[t]
  \begin{center}
    \includegraphics[width= \columnwidth]{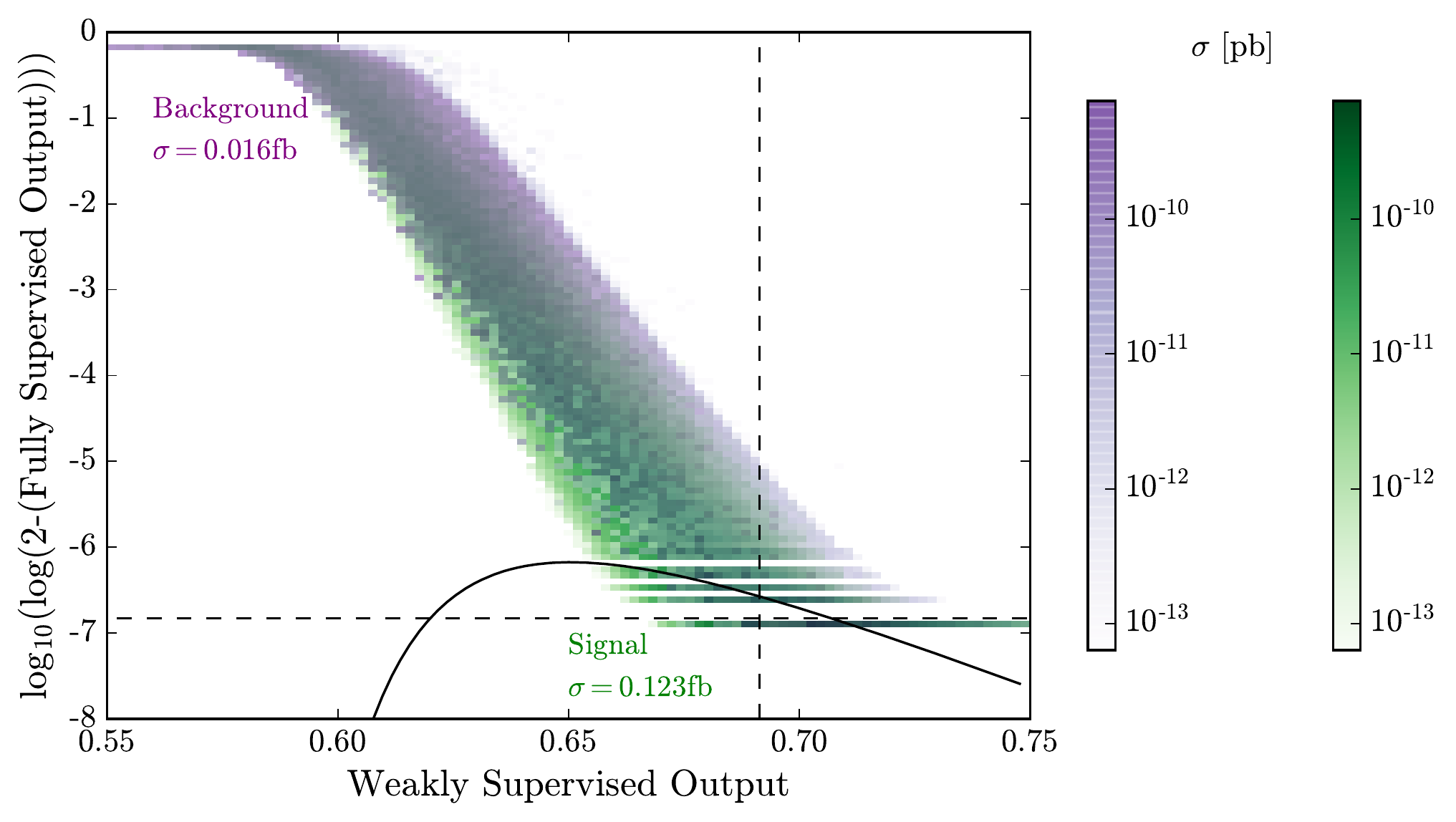}
	\caption{Two-dimensional histogram of the weighted $Z+$jets background and gluino pair production signal events. The horizontal axis corresponds to the output of the weakly supervised network, with the background peaked towards values of 0.6 and signal peaked towards 0.7. The vertical axis is the output for each event from the fully supervised network, transformed to separate the values close to 1. The larger, negative values correspond to signal, while close to 0 is the background. The two dashed lines show the values of the cuts places on either individual network. The solid line represents the cut found by training a fully supervised network which uses only these two axes as input. This cut reduces the background (for the same signal cross section) by \SI{15.8}{\percent} from the best cut on either network alone.}
	\label{fig:BothOutputsBSM}
  \end{center}
\end{figure}

To demonstrate this, \cref{fig:BothOutputsBSM} provides a 2D histogram for the weighted signal (green) and background (purple) events. The output of the weakly supervised network for each event lies along the horizontal axis. The fully supervised network places the signal events very close to 1. Therefore, we transform the output of the fully supervised network by taking $\log(2-\text{output})$, so that events close to 1 are given by a small number, and events close to zero are mapped to $\log 2$. The outputs are then spread further by taking the $\log_{10}$, as shown on the vertical axis. The dashed lines show the (transformed) cuts used at the end of \cref{sec:BSM_initial}, with the signal region being to the right or bottom for the weakly and fully supervised networks, respectively. 

A new, fully supervised network with 30 nodes on the hidden layer is trained only using the outputs of the previous networks as input. The network is trained using the \textsc{Adam} optimizer with a learning rate of 0.001 for 50 epochs using all of the weighted background and signal events. The events were first re-weighted so that the summed weights for all of the background is only 5 times that of summed signal events. With this network, we then choose a value of cut which gives a similar signal cross section as used before. The remaining background cross section after the cut is \SI{0.016}{\fb}. This is a \SI{15.8}{\percent} reduction in the background compared to the fully supervised network alone. Furthermore, it appears that the lower edge of the signal blob also does not contain background, so one could add in these extra regions by hand for an additional improvement. While the idea of ensembleing or stacking machine learning models together to boost performance is not new~\cite{1688199,Rokach2010, DBLP}, this example shows that it can be done simply using both fully and weakly supervised networks.

Machine learning as applied to LHC is still in a rudimentary phase.  As these tools become more relevant, it is crucial that systematic studies like the one presented here are done.  Clearly weakly supervised neural networks are extremely powerful, and we look forward to seeing their application to a variety of physical processes in the future.

\acknowledgments

We thank Matthew Dolan, Sebastian Macaluso, Eric Metodiev, Ben Nachman, Francesco Rubbo, David Shih, and Jesse Thaler for useful comments.  TC is supported by an LHC Theory Initiative Postdoctoral Fellowship, under the National Science Foundation grant PHY-0969510, and by the U.S. Department of Energy (DOE), under grant number DE-SC0018191.  MF and BO are supported by the DOE under contract DE-SC0011640.

\appendix
\section*{Appendix}
\section{Choice of Loss Function}
\label{Sec:Appendix}

Through much of this paper, we have compared fully and weakly supervised networks to argue that there is negligible performance loss when using weak supervision. However, the analytic arguments made here, as well as in~\cite{Dery:2017fap, Metodiev:2017vrx}, imply that classifying on label proportions should work just as well as classifying with event-by-event labels.  These proofs work in the limit of infinite statistics such that the  optimal classifier has been obtained, and as such, they do not depend on how this has been achieved.  Therefore, it is important to understand how these statements apply in practice, where the choice of hyper-parameters can impact the performance.  It is for this reason that we chose the same hyper-parameters and optimizers when making our numerical comparisons between different supervision techniques.  In practice, when applying a network to a physics scenario, a careful tuning of the hyper-parameters should be performed; it is likely that the optimized weakly supervised network would use different choices than for a fully supervised network.

\begin{figure}[tb!]
  \begin{center}
    \includegraphics[width= 0.5\columnwidth]{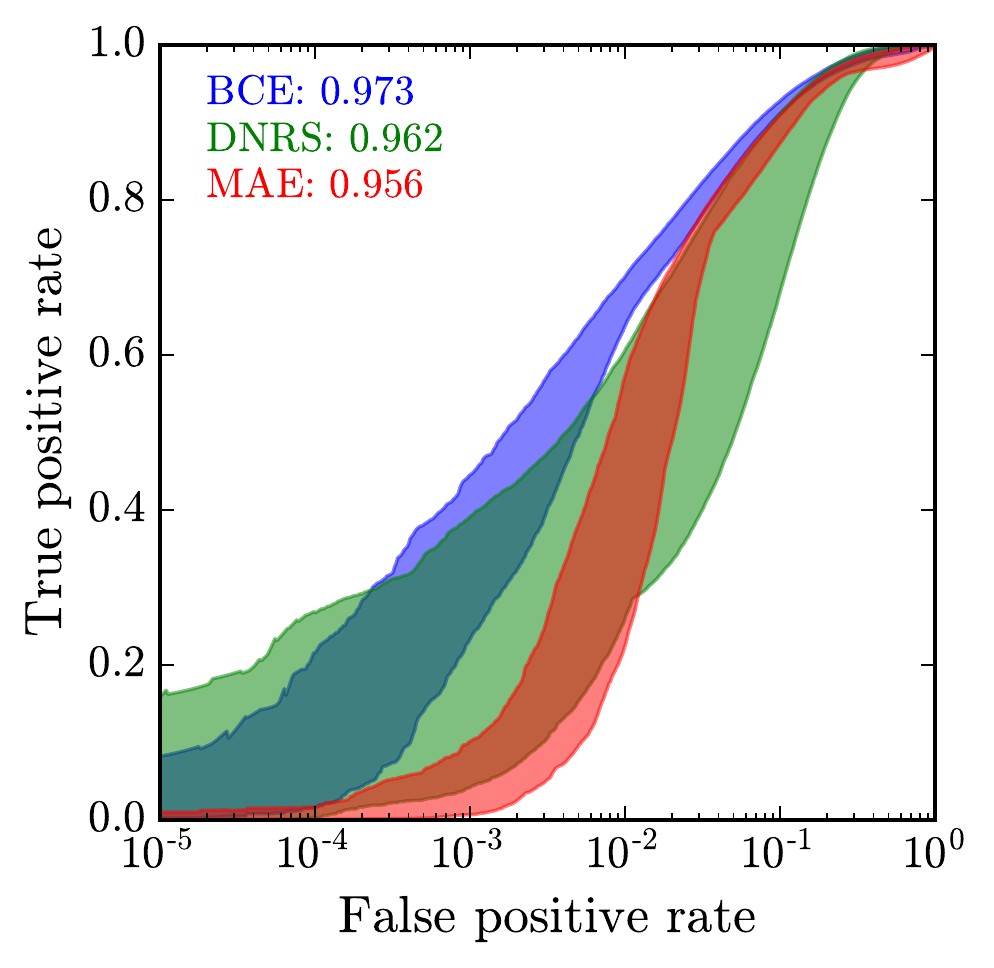}
	\caption{The spread in the ROC curves for 100 different trainings of weak supervision for different choices of the loss function. The mean area under the curve is denoted in the legend. The ordering of the legened matches the curves at a false positive rate of $10^{-3}$.}
	\label{fig:LossFunctionComparison}
  \end{center}
\end{figure}

To emphasize this point, we show how the choice of loss function impacts the ability of the network to learn to classify.  Using our toy model defined in \cref{sec:toys}, we train 100 independent weakly supervised networks.  The fraction labels are chosen to be the truth values of 0.4 and 0.7. The first loss function is binary cross entropy, which is what is used throughout this paper since it yields the strongest discriminating power. Additionally, we tested the Mean Absolute Error (MAE), given by
\begin{equation}
\label{eqn:MAE}
  \ell_\text{MAE}(\{ f_t \}, \{ y_p \})
    = \sum_i \big| f_{t,i} - y_{p,i} \big|,
\end{equation}
and the original loss function proposed in \cite{Dery:2017fap},
\begin{equation}
\label{eqn:DNRS}
  \ell_\text{DNRS}(\{ f_t \}, \{ y_p \})
    = \big| \avg{f_{t,i}} - \avg{y_{p,i}} \big|,
\end{equation}
where the angled brackets denote the mean of a particular batch of events. The networks are optimized using \textsc{Adam} with a learning rate of 0.0015 training on 100 epochs with batch sizes of 32. The spread in the obtained ROC curves is shown in \cref{fig:LossFunctionComparison}, along with the average AUC. Both finite statistics and differing convergence rates contribute to the performance differences observed. The DNRS loss function shows slower convergence (and larger variance) when training due to the need to infer implied distributions for the signal and background, while the differences between the BCE and MAE loss functions are due largely to finite statistics, with the former giving greater weight to errors at high purity. All of this emphasizes the need to optimize hyperparameters when choosing a particular strategy.

As measured by the AUC, each of the different choices of loss function yields near optimal performance. However, their behavior diverges at small false positive rates.  From the analytic arguments, weak supervision should perform as well as full supervision with optimal learning algorithms and infinite statistics. These results show that in practice, the method used to obtain the classifier impacts how closely it approximates the optimal classifier. Training details and choice of initialization can help minimize the variance given a particular training set, but the extent to which such systematics will be important in practice will depend on the specifics, such as choice of working point, of how the classifier is used.

\bibliographystyle{utphys}
\bibliography{WeakBib}

\end{document}